\documentclass[aps,pre,twocolumn,groupedaddress,amsmath,showpacs]{revtex4}
\usepackage{graphics}

\begin{document}

\title{A generalized thermodynamics for power-law statistics}

\author{Massimo Marino}
\email[]{Massimo.Marino@unimi.it}
\affiliation{Dipartimento di Matematica, Universit\`{a} di Milano,
via Saldini 50, I-20133 Milano (Italy)}

\date{September 17, 2007}

\begin{abstract}
We show that there exists a natural way to define a condition of
generalized thermal equilibrium between systems governed by
Tsallis thermostatistics, under the hypotheses that i) the
coupling between the systems is weak, ii) the structure functions
of the systems have a power-law dependence on the energy. It is
found that the $q$ values of two such systems at equilibrium must
satisfy a relationship involving the respective numbers of degrees
of freedom. The physical properties of a Tsallis distribution can
be conveniently characterized by a new parameter $\eta$ which can
vary between $0$ and $+\infty$, these limits corresponding
respectively to the two opposite situations of a microcanonical
distribution and of a distribution with a predominant power-tail
at high energies. We prove that the statistical expression of the
thermodynamic functions is univocally determined by the
requirements that a) systems at thermal equilibrium have the same
temperature, b) the definitions of temperature and entropy are
consistent with the second law of thermodynamics. We find that,
for systems satisfying the hypotheses i) and ii) specified above,
the thermodynamic entropy is given by R\'enyi entropy.
\end{abstract}


\maketitle

\section{Introduction}

In recent years a considerable amount of theoretical research
activity \cite{abeoka,tsallisbr,gellmann,erice,creta,biblio} has
been stimulated by the conjecture that a certain class of
dynamical systems might be governed by a peculiar thermal
statistics, in which the usual Boltzmann--Gibbs exponential
distribution is replaced by a type of power-law distribution
originally proposed by Tsallis \cite{tsallis88}. Such a conjecture
seems to be supported by phenomenological and computational
evidence coming from various domains of physics and other
disciplines.

It has been proposed in various ways to define a generalized
entropy function, according to formulae involving a parameter $q$
directly related to the characteristic exponent of the power-law
distribution (whence the frequently used name of
``$q$-distributions''). It appears then natural to extend to these
systems also the concept of temperature, in such a way that the
validity of the second law of thermodynamics is preserved. However
some difficulties have to be faced in order to define the
condition of thermal equilibrium and to consistently generalize
for these systems the zeroth law of thermodynamics. The simplest
possibility is apparently that of treating systems at thermal
equilibrium as statistically independent of one another
\cite{abe}. However, it has already been recognized that the
assumption of such an independence has no fundamental
justification. Furthermore, the probability distribution for a
composed system obtained by multiplication of two power-law
distributions does not belong in general to the same category.
This problem is not present in the Boltzmann--Gibbs case thanks to
the identity $e^{-\beta E_1}e^{-\beta E_2}= e^{-\beta(E_1+E_2)}$,
which represents a peculiar property of the exponential function.
As a way out of this difficulty, a generalization of the zeroth
law in the spirit of superstatistics \cite{beck} has recently been
proposed \cite{vignat}.

A different approach to the thermodynamics of systems with
power-law statistics has been developed following the observation
that a distribution of this type can also be formally derived
using standard arguments of equilibrium thermodynamics. In fact,
in the same way as the usual Boltzmann--Gibbs distribution is
derived for a system in thermal contact with an infinite heat
bath, a power-law distribution is obtained when the heat bath is
instead characterized by a finite number of degrees of freedom and
a power-law density-of-states function \cite{plastino,almeida1,
adib}. In such a situation, it becomes natural to define the
physical temperature by making use of the equipartition theorem
\cite{almpot}. It has been shown that, when the inverse of this
temperature is used as the integrating factor of the exchanged
heat, in accordance with the second law of thermodynamics, the
definition of the entropy has to be modified with respect to the
original Tsallis formula \cite{potiguar,almeida}. It has however
to be noted that the presence of a finite heat bath can only be
assumed when the power-law distributions present a finite energy
cut-off, which is not the case for all the possible ranges of the
parameter $q$.

The distinction between ``observed system'' on the one side, and
``heat bath'' on the other, which has been followed in Refs.\
\cite{plastino,almeida1,adib, potiguar,almeida} in analogy with
familiar considerations of standard thermodynamics, appears
somewhat fictitious when both systems are supposed to be finite.
In the present paper we shall then abandon the concept of heat
bath altogether, and we shall only use that of mutual thermal
(meta)equilibrium, in a form which can in principle be applied to
systems obeying any (\emph{a priori} arbitrary) generalized type
of thermal statistics. If we consider a system composed of parts
in mutual thermal contact, it is obvious that when the system is
in a stationary state, its parts are in thermal metaequilibrium
with one another (of course, the parts in question may be either
of homogeneous or of heterogeneous physical nature). We shall
investigate, with the widest possible generality, the logical
consequences of this simple fact in connection with the laws of
thermodynamics. In particular, in order to determine the correct
definition for the absolute temperature, we shall naturally impose
the condition that, in accordance with the zeroth law, systems at
equilibrium must have the same temperature. We shall see that such
an approach applies equally well to Tsallis distributions having a
finite energy cutoff, and to those showing instead a power tail
extending to infinitely large energies.

From a mathematical point of view, in order to apply in a
consistent way the concept of thermal equilibrium to states
described by Tsallis statistics, we have first of all to establish
under which conditions the probability distribution for a composed
system, and the marginal distributions which are derived from it,
can all have simultaneously a power-law dependence on the energy.
We will show that a sufficient condition is that also the
densities-of-states of all the subsystems considered are given by
power functions of the energy. Of course we know that this is
indeed the case for ideal gases, and represents a good
approximation for a quite large class of systems. For these
systems we can then interpret Tsallis statistics as a general
description of metaequilibrium states, which is consistent with
the zeroth law of thermodynamics. With respect to standard
statistical mechanics \cite{gibbs,khin}, which is only based on
the Boltzmann--Gibbs distribution, the new theory contains a set
of additional parameters, the $q$ values of the systems
considered. It will be shown that these parameters, in order for
thermal equilibrium to establish, must be in a certain
relationship with the respective numbers of degrees of freedom.

The first and second laws of thermodynamics maintain their
validity for systems governed by power-law distributions. An
important point, which will emerge from our analysis, is that
these laws, when considered in combination with the zeroth one,
univocally determine the expression of both the temperature and
entropy functions. It will turn out that, for systems to which our
starting hypotheses can be applied, the correct expression for the
entropy coincides with that originally proposed by R\'enyi
\cite{renyi,renyi2}.

The consideration of power-law densities of states will also allow
us to clarify some aspects of power-law distributions, which have
often been overlooked in the past literature. It will be shown
that the distributions with a finite energy cutoff have properties
intermediate between the microcanonical and the canonical
ensembles (which are exactly obtained in the two limiting cases
$q=0$ and $q=1$ respectively), whereas those with an infinite
power tail represent a basically new physical situation. The
consideration of the parameter $q$ alone can be misleading. In the
case of large systems, $q$ is in fact always necessarily close to
1, but we will show that this by no means implies that the
character of the distribution is necessarily close to
Boltzmann--Gibbs. We propose the introduction of a new parameter
$\eta$, expressed as a function of both $q$ and the number of
degrees of freedom, which provides an estimate of the actual
deviations from the canonical distribution.

Sections \ref{prelim0}--\ref{pdof} deal with some basic facts
about statistical thermodynamics, power-law distributions and
power-law densities of states respectively. Although some of the
results we shall obtain in these sections are not completely new,
we shall present them with an emphasis on those aspects which are
at the basis of our particular approach to generalized
thermodynamics. This will lead us in Sec.\ \ref{class} to our new
classification of power-law distributions by means of the
parameter $\eta$, and in Sec.\ \ref{thermo} to the main results
about the statistical expression of thermodynamic temperature and
entropy.

\section{Preliminaries} \label{prelim0}

\subsection{Temperature and thermal equilibrium} \label{prelim}

Let us consider a system $\cal S$ composed of two weakly coupled
subsystems ${\cal S}_1$ and ${\cal S}_2$. We shortly denote with
$z_1$ the set of $2\nu_1$ canonical variables $q_{1,1},\ldots,
q_{1,\nu_1}, p_{1,1},\ldots, p_{1,\nu_1}$ of the phase space
$\Gamma_1$ of the subsystem ${\cal S}_1$, and with $H_1(z_1)$ the
corresponding hamiltonian function. If the same notation is used
for the subsystem ${\cal S}_2$, the fact that the coupling between
the two is weak allows us to express with good approximation the
hamiltonian $H$ of the complete system $\cal S$ as $H\simeq
H_1(z_1)+H_2(z_2)$.

We define the function $\Omega_1(E_1)$ as the volume of the region
of space $\Gamma_1$ corresponding to energies of the subsystem
${\cal S}_1$ lower than $E_1$. Introducing the step function
$\theta$ such that $\theta(x)=1$ for $x\geq 0$, $\theta(x)=0$ for
$x<0$, we can write
\begin{equation}
\Omega_1(E_1)=\int_{\Gamma_1}dz_1\,\theta(E_1-H_1(z_1))\ ,
\end{equation}
where $dz_1$ stands for a $2\nu_1$-dimensional
volume element of $\Gamma_1$. The derivative $\omega_1(E_1)$ of
the above function can then be identified with the ``density of
states'' of ${\cal S}_1$:
\[
\omega_1(E_1)= \Omega'_1(E_1)
=\int_{\Gamma_1}dz_1\,\delta(E_1-H_1(z_1))\ .
\]
After defining in the same way the functions $\Omega_2$ and
$\omega_2$ for the subsystem ${\cal S}_2$, we obtain
\cite{khin,gibbs} for the corresponding functions of the complete
system $\cal S$:
\begin{eqnarray}
\Omega(E)&=&\int_{\Gamma}dz_1dz_2\,\theta(E-H_1(z_1)- H_2(z_2))
\nonumber \\
&=&\int dE_1 \,\omega_1(E_1)\Omega_2(E-E_1) \nonumber \\
&=&\int dE_2 \,\Omega_1(E-E_2)\omega_2(E_2) \ . \label{omega}
\end{eqnarray}

With each macroscopic state of the system is associated a
probability density function $\rho$ on the space $\Gamma$. As a
consequence of Liouville's theorem \cite{khin}, a stationary
probability density can only be a function of constants of motion
of the system. It is usual in statistical mechanics to assume that
$\rho$ is actually only a function of the energy. We will then
write $\rho(z_1,z_2)=\Phi(H_1(z_1)+H_2(z_2))$, where $\Phi$ is a
function of a single real variable, on which no particular
hypothesis is made for the moment. The marginal distribution
$\rho_1$ for the subsystem ${\cal S}_1$, defined as
$\rho_1(z_1)=\int_{\Gamma_2} dz_2\,\rho(z_1,z_2)$, can then be
expressed as $\rho_1(z_1)=\Phi_1(H_1(z_1))$, where
\begin{subequations} \label{marginal}
\begin{eqnarray}
\Phi_1(E_1)&=&\int_{\Gamma_2} dz_2\,\Phi(E_1+H_2(z_2)) \nonumber \\
&=& \int dE_2\,\omega_2(E_2)\Phi(E_1+E_2) \ .\label{margin1}
\end{eqnarray}
We have similarly $\rho_2(z_2)=\Phi_2(H_2(z_2))$, with
\begin{equation}
\Phi_2(E_2)=\int dE_1\,\omega_1(E_1)\Phi(E_1+E_2)\ .
\label{margin2}
\end{equation} \end{subequations}

Since $\omega(E)\Phi(E)$ represents the probability density
function for the energy, defining the functional $\Theta$ as the
average of the function $\Omega(E)/\omega(E)$ we have
\begin{equation}
\Theta =\int dE\,\Phi(E)\Omega(E) \label{beta0}
\end{equation}
and similarly for $\Theta_1$ and $\Theta_2$. Making use of Eqs.\
(\ref{omega}) and (\ref{marginal}) we find
\begin{eqnarray*}
\int dE\,\Phi(E)\Omega(E) &=&\int dE_1\,\Phi_1(E_1)\Omega_1(E_1) \\
&=&\int dE_2\,\Phi_2(E_2)\Omega_2(E_2)\ ,
\end{eqnarray*}
or equivalently \cite{gibbs}
\begin{equation}
\Theta=\Theta_1=\Theta_2\ . \label{theta}
\end{equation}
For the case of a system ${\cal S}$ described by a microcanonical
ensemble, the quantity $\Theta$ was given the name of ``empirical
temperature'' in Ref.\ \cite{munster}, on the grounds of the
equipartition theorem. For our present purposes, the main point is
that Eq.\ (\ref{theta}) has here been proved in a completely
general way, independently of any assumptions about the specific
form of the probability density or of the density-of-states
functions. As we explained in the Introduction, we assume that the
parts of an isolated system in a stationary state are in thermal
(meta)equilibrium with each other. We can then affirm that the
functional $\Theta$ assumes the same value for any two systems for
which this equilibrium condition is fulfilled. Such a result is
not sufficient to conclude at this point that $\Theta$ represents
the correct generalized definition of thermodynamical temperature,
since the second law has not yet been taken into consideration.
Besides, there could exist in principle other quantities
independent of $\Theta$ which, at least for a certain class of
systems, enjoy a property of the same form as that expressed by
Eq.\ (\ref{theta}). If we denote the set of such hypothetical
quantities with $r$, we can provisionally conclude that a
generalized formula for the inverse temperature $\beta$ in
statistical mechanics, in order to be consistent with the zeroth
law of thermodynamics, has to be expressible as an appropriate
function of $\Theta$ and $r$ of the form
\begin{equation}
\beta=f_r(\Theta)\ . \label{beta}
\end{equation}

\subsection{Temperature and thermodynamics} \label{tethe}

If the hamiltonian $H$ depends, besides on $z$, also on some
external parameters $a_j$ (such as for instance the volume, or the
linear dimensions of the system's container), then under the
infinitesimal variations $da_j$ of these parameters the system
performs on the surrounding environment the average work
\begin{equation}
\delta W_{\rm rev}=-\int_\Gamma dz\,\rho dH
\end{equation}
with $dH= \sum_j (\partial H/\partial a_j)da_j$. From the first
law of thermodynamics we get
\begin{eqnarray}
\delta Q_{\rm rev}&=& dU+\delta W_{\rm rev}  \nonumber \\
&=&d\left(\int_\Gamma dz\,\rho H \right)- \int_\Gamma dz\,\rho dH
\nonumber \\
&=& \int_\Gamma dz\,\delta\rho H\ . \label{qrev}
\end{eqnarray}
The above equation provides the expression in the language of
statistical mechanics for the quantity of heat $\delta Q_{\rm
rev}$ which is exchanged by the system in an infinitesimal
reversible transformation.

According to the so called Clausius theorem, which is based on the
second law of thermodynamics, $\delta Q_{\rm rev}$ admits an
integrating factor which can be identified with the inverse of the
absolute temperature. It can be easily shown that a sufficient
condition for the validity of an equivalent theorem in statistical
thermodynamics is the existence of a real function $F$, with a
concavity of definite sign, such that the probability distribution
$\rho$ can be obtained as a stationary point of the functional
\begin{equation}
\tilde S=\int_\Gamma dz\,F(\rho) \label{entrof}
\end{equation}
under the constraints
\begin{subequations} \label{constra}
\begin{equation}
\int_\Gamma dz\,\rho=1 \label{constra1}
\end{equation}
and
\begin{equation}
\int_\Gamma dz\,\rho H=U\ . \label{constra2}
\end{equation}
\end{subequations}
Equation (\ref{constra1}) simply imposes the correct normalization
for the probability density, while Eq.\ (\ref{constra2}) fixes the
value $U$ for the mean energy of the system. Introducing the
Lagrange multipliers $\tilde\alpha$ and $\tilde\beta$, such a
constrained extreme can be located as the zero of a functional
derivative in the following way:
\begin{eqnarray}
0&=&\frac{\delta}{\delta \rho(z)}\left(\tilde S -\tilde\alpha
\int_\Gamma dz\,\rho -\tilde\beta \int_\Gamma dz\,\rho H \right)
\nonumber \\
&=&F'(\rho(z))-\tilde\alpha-\tilde\beta H(z)\ . \label{moltiplic}
\end{eqnarray}
The hypothesis we have made on the concavity of $F$ guarantees
that $F'$ is a monotone function, and therefore admits an inverse
function which we call $G$. From the last equation it then follows
that we can write $\rho(z)=\Phi(H(z))$ with
\begin{equation}
\Phi(E)=G(\tilde\alpha+\tilde\beta E)\ . \label{staz}
\end{equation}
Equation (\ref{moltiplic}) implies that
\[
d\left(\tilde S -\tilde\alpha \int_\Gamma dz\,\rho -\tilde\beta
\int_\Gamma dz\,\rho H \right)=0
\]
for any arbitrary variations $\delta\rho(z)$ at fixed $a_j$. It
follows
\begin{equation}
d\tilde S = \tilde\alpha d\left(\int_\Gamma dz\,\rho\right)
+\tilde\beta \int_\Gamma dz\,\delta\rho H = \tilde\beta\delta
Q_{\rm rev}\ ,
\end{equation}
where for the last step we have used Eq.\ (\ref{qrev}) and the
fact that $d(\int_\Gamma dz\,\rho)=0$ on account of Eq.\
(\ref{constra1}). The above equation shows that the state function
$\tilde\beta$, defined as the Lagrange multiplier which solves the
system of equations (\ref{constra})--(\ref{moltiplic}), is an
integrating factor for $\delta Q_{\rm rev}$.

Such an integrating factor is in principle not unique. Let us
define
\begin{equation}
S=kg(\tilde S)\ , \label{esse}
\end{equation}
where $k$ is the Boltzmann constant and $g$ any monotone function.
We then have, denoting with $g'$ the derivative of $g$,
\begin{equation}
dS = kg'(\tilde S)d\tilde S = k\beta \delta Q_{\rm rev}\ ,
\label{desse}
\end{equation}
with
\begin{equation}
\beta=\tilde\beta g'(\tilde S)\ . \label{betaii}
\end{equation}
Equation (\ref{desse}) shows that also the new state function
$\beta$, defined by Eq.\ (\ref{betaii}), is an integrating factor
for $\delta Q_{\rm rev}$. The analytical expressions of the
functions $f$ and $g$ appearing respectively in Eqs.\ (\ref{beta})
and (\ref{esse}) are up to this point undetermined: they are
related to each other by the condition that they must provide the
same value for $\beta$.

\subsection{The Boltzmann--Gibbs distribution}

If we put in Eq.\ (\ref{entrof}) $F(\rho)=-\rho\log(h^\nu\rho)$,
where $h$ is a constant with the dimensions of an action and
$2\nu$ is the dimension of the phase space $\Gamma$, we obtain
from Eq.\ (\ref{staz}) $\Phi(E)=Z^{-1}e^{-\tilde\beta E}$, where
the partition function $Z=h^\nu e^{1+\tilde\alpha}$ provides the
required normalization factor for the probability density. From
Eq.\ (\ref{beta0}) we obtain \cite{gibbs}
\begin{eqnarray}
\Theta&=&\frac 1 Z\int_{E_{\rm min}}^{+\infty} dE\,e^{-\tilde
\beta E}\Omega(E) \nonumber \\
&=&\frac 1{Z\tilde\beta} \int_{E_{\rm min}}^{+\infty}
dE\,\omega(E)e^{-\tilde\beta E} =\frac 1{\tilde\beta}
\label{beta0bg}
\end{eqnarray}
where, integrating by parts, we have used the fact that
$\Omega(E_{\rm min})=0$. Hence, according to the considerations of
the previous subsection we can simply put $f_r(\Theta)=1/ \Theta$
in Eq.\ (\ref{beta}) and $g(\tilde S)=\tilde S$ in Eq.\
(\ref{betaii}), obtaining
\begin{equation}
\tilde\beta =1/\Theta =\beta \ . \label{bbb}
\end{equation}
Furthermore, for an ideal gas composed of a large number $N$ of
molecules, it is found that $\beta = PV/N$, where $P$ is the
pressure and $V$ the volume occupied by the gas. Therefore, on the
basis of the laws of thermodynamics and of the state equation of
perfect gases, one can rightfully associate with the
Boltzmann--Gibbs distribution
\begin{equation}
\Phi(E)=Z^{-1}e^{-\beta E} \label{phibg}
\end{equation}
the absolute temperature $T=1/k\beta$ and the entropy
\begin{equation} \label{essebg}
S_{BG}=-k\int_\Gamma dz\,\rho\log(h^\nu\rho)\ .
\end{equation}

It has been assumed until now that the system can be correctly
described by classical mechanics. However, if we identify $h$ with
the Planck constant, according to Heisenberg uncertainty principle
we can associate with each element of $\Gamma$ of volume $h^\nu$ a
quantum state with occupancy probability $P=h^\nu \rho$. On the
basis of this correspondence we can rewrite the Boltzmann--Gibbs
entropy (\ref{essebg}) in the form
\begin{equation}
S_{BG}=-k\sum_i P_i \log P_i \ ,\label{sbg}
\end{equation}
the sum being extended to a complete set of quantum states of the
system.

\section{Power-law distribution functions} \label{pldf}

\subsection{Definition} \label{defin}

Let us fix the arbitrary additive constant of the hamiltonian
function in such a way that its minimum value is zero. We refer as
``power-law distributions'' to stationary probability density
functions $\rho(z)=\Phi(H(z))$ for which the function $\Phi$,
defined on the interval $[0,+\infty)$, takes one of the two
alternative forms
\begin{subequations}
\begin{equation}
\Phi(E)=D^{-1}(E-E_0)^{-p} \quad \text{for} \quad 0\leq E< +\infty
\label{piu}
\end{equation}
with $E_0<0$, or
\begin{equation}
\Phi(E)=\begin{cases}D^{-1}(E_0-E)^{-p}\quad & \text{for} \quad
0\leq E < E_0 \\
0 \quad & \text{for}\quad E\geq E_0 \ .
\end{cases} \label{meno}
\end{equation}
\end{subequations}
In the above equations $p$ is a real parameter, and $D$ a
normalization coefficient such that
\begin{equation}
\int_0^{+\infty} dE\,\omega(E)\Phi(E)=1\ . \label{probab}
\end{equation}
In Eq.\ (\ref{meno}) the usual Tsallis cut-off prescription
\cite{tsallis88,curado} has been adopted. It has become customary
in the literature to equivalently express these distributions by
means of the $q$-exponential function \cite{tsallisbr}, but this
would not here be helpful in view of the formal manipulations we
shall need to perform.

The requirement that the integral on the left-hand side of the
last equation be finite imposes limitations on the possible values
of $p$. In the case of Eq.\ (\ref{piu}), if we assume that the
behavior of the density-of-states function in the limit of high
energies is expressible as
\begin{equation}
\omega(E)\propto E^{s-1}\quad \text {for}\quad E\rightarrow
+\infty\ , \label{asintot}
\end{equation}
with $s>0$, then a necessary condition is $p>s$. Furthermore, if
the first law of thermodynamics has to hold, it must be possible
to define the average energy of the system as
\begin{equation}
U=\int_0^{+\infty} dE\,\omega(E)\Phi(E)E\ . \label{energia}
\end{equation}
Therefore, in order for $U$ to be finite, in the case of Eq.\
(\ref{piu}) it is actually necessary that $p>s+1$. In the case
instead of Eq.\ (\ref{meno}), the requirement that both the
integrals in Eqs.\ (\ref{probab}) and (\ref{energia}) be
convergent in a left neighborhood of $E_0$ is equivalent to the
condition $p<1$.

In the rest of the present paper we shall often write the
power-law distributions in the general form
\begin{equation}
\Phi(E)=D^{-1}|E-E_0|^{-p}\ , \label{phi}
\end{equation}
where it is understood that $E$ can only take values such that
$E-E_0>0$ (resp. $E-E_0<0$) whenever $p>1$ (resp. $p<1$). The
qualitative behavior of the function $\Phi$ in the four cases
$p>1$, $0<p<1$, $-1<p<0$ and $p<-1$ is displayed for the sake of
clarity in Fig.\ 1. When Eq.\ (\ref{asintot}) holds, for the
reasons we have explained $p$ can never lie within the interval
$[1,1+s]$.

\begin{figure}
\includegraphics{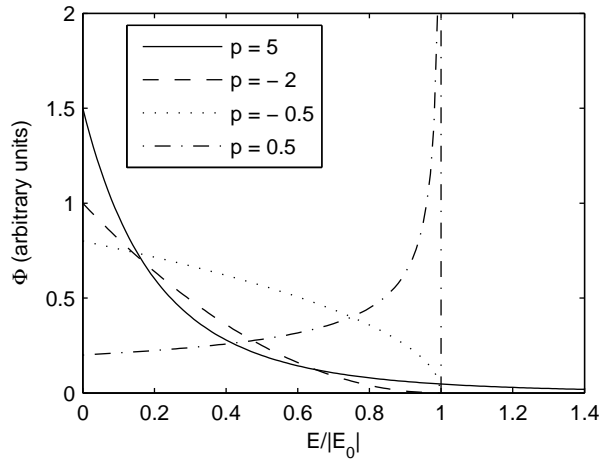}
\caption{Qualitative behavior of the function $\Phi$ for different
values of $p$.}
\end{figure}

\subsection{$q$-Distributions and thermodynamics} \label{qtherm}

It is easy to see that a power-law distribution can be obtained
via a maximization procedure like that illustrated in Sec.\
\ref{tethe}. In fact, if in Eq.\ (\ref{entrof}) we take $F(\rho)=
h^{\nu(q-1)} \rho^q$, where $q$ is a real number, then Eq.\
(\ref{staz}) provides
\begin{equation}
\Phi(E)=h^{-\nu}\left(\frac{\tilde\alpha +\tilde\beta
E}q\right)^{1/(q-1)} \ , \label{stazq}
\end{equation}
which coincides with Eq.\ (\ref{phi}) when putting
\begin{eqnarray}
p&=&\frac 1{1-q} \label{pq} \\
E_0&=&-\tilde\alpha/\tilde\beta \\
D&=& h^{\nu}|\tilde\beta/q |^{1/(1-q)}\ . \label{di}
\end{eqnarray}
Looking at Eq.\ (\ref{stazq}) one also sees that $\Phi$ will be of
the form (\ref{piu}) when $\tilde\beta/q>0$, and we have shown
that in such a case the existence of a finite mean energy requires
$p>1$ as a necessary condition. Viceversa, $\Phi$ will be of the
form (\ref{meno}) when $\tilde\beta/q<0$, and in that case we must
then have $p<1$.

We can rewrite Eq.\ (\ref{entrof}) in the form
\begin{equation}
\tilde S_q=h^{-\nu}\int_\Gamma dz\, (h^\nu\rho)^q \ ,
\label{entroq}
\end{equation}
which in terms of the discrete probabilities of the quantum states
becomes
\begin{equation}
\tilde{S}_q=\sum_i P_i^q\ . \label{tildes}
\end{equation}
According to Eq.\ (\ref{esse}) the entropy will be of the general
form
\begin{equation}
S_q=k g_q(\tilde S_q)\ , \label{esseq}
\end{equation}
where $g_q$ is an appropriate monotone function which may contain
$q$ as a parameter. Tsallis originally derived his power-law
distribution from the function \cite{tsallis88}
\begin{equation}
S^T_{q} = k\frac{1- \tilde S_q}{q-1}\ . \label{tsallis}
\end{equation}
On the other hand, the expression
\begin{equation}
S^R_q=k\frac{\log \tilde S_q}{1-q} \label{renyi}
\end{equation}
had already been considered by R\'enyi in the context of
information theory \cite{renyi}. Other variants have also
occasionally received some attention in the literature: we may
here mention Tsallis normalized entropy \cite{lands}
\begin{equation}
S^N_{q} = k\frac{1- \tilde S_q}{(q-1)\tilde S_q} \label{normal}
\end{equation}
and Tsallis ``escort'' entropy \cite{tsescort}
\begin{equation}
S^E_{q}=kq\frac{1- (\tilde S_q)^{-1/q}}{1-q}\ . \label{escort}
\end{equation}
Note that $P_i^q \sim P_i+(q-1)P_i\log P_i$ for $q \rightarrow 1$,
whence $\tilde S_q \sim 1+(q-1)\sum_i P_i \log P_i$. It is then
easy to see that all the four functions given in Eqs.\
(\ref{tsallis})--(\ref{escort}) tend to the Boltzmann--Gibbs
entropy (\ref{sbg}) for $q \rightarrow 1$.

From the considerations made in Sec.\ \ref{tethe} it is clear that
all these functions (and other possible ones) deserve equally well
the name of entropy as far as the second law of thermodynamics is
concerned, since for each of them one can define a state function
$\beta$ such that Eq.\ (\ref{desse}) is satisfied. The criterium
for the correct choice is provided in fact by the zeroth law. In
the following of the present paper, after a closer analysis of the
condition of thermal equilibrium between power-law distributions,
we shall aim at determining the analytic form of $g_q$ for which
$\beta$ can be expressed in the form (\ref{beta}), with $\Theta$
given by Eq.\ (\ref{beta0}).

\subsection{A generalized partition function}

When a system is described by a power-law distribution of the form
(\ref{phi}), some of the quantities previously introduced can be
conveniently expressed with the aid of the auxiliary function
\begin{eqnarray}
\zeta(E_0)&=&\int_I dE\,\omega(E)|E-E_0|^{1-p} \nonumber \\
&=& |1-p|\int_I dE\,\Omega(E)|E-E_0|^{-p} \label{zeta}
\end{eqnarray}
which, in the same way as an analogous one introduced in Ref.\
\cite{potiguar}, plays the role of a generalized partition
function for this class of systems. According to the
considerations made in Sec.\ \ref{defin}, when $p>1$ the function
$\zeta$ is defined for $E_0<0$ and the integration domain $I$ in
Eq.\ (\ref{zeta}) is the interval $[0,+\infty)$, whereas when
$p<1$ we have $E_0>0$ and $I=[0,E_0)$.

From Eq.\ (\ref{probab}) we get
\begin{equation}\label{prob2}
    D(E_0)=\int_I dE\,\omega(E)|E-E_0|^{-p} = \frac {\zeta'
    (E_0)} {|1-p|}  \ .
\end{equation}
We can then express the function $\Theta$ defined by Eq.\
(\ref{beta0}) as
\begin{eqnarray}\label{beta0z}
    \Theta &=& \frac 1{D(E_0)}\int_I dE\,\Omega(E)
    |E-E_0|^{-p} \nonumber \\
    &=&\frac{\zeta(E_0)}{|1-p|D(E_0)} = \left[\frac d{dE_0}\log
    \zeta(E_0)\right]^{-1} \ .
\end{eqnarray}

We can also evaluate the mean energy $U$ using the relation
\begin{eqnarray}
|U-E_0| &=& \int_I dE\,\omega(E)\Phi(E)|E-E_0| \nonumber \\
&=& \frac 1{D(E_0)}\int_I dE\, \omega(E)|E-E_0|^{1-p} \nonumber \\
&=& \frac{\zeta(E_0)}{D(E_0)} 
= |1-p| \Theta \ . \label{uz}
\end{eqnarray}
Since it is obviously $U>E_0$ (resp. $U<E_0$) for $p>1$ (resp.
$p<1$), Eq.\ (\ref{uz}) is equivalent to
\begin{equation}
U=E_0+(p-1)\Theta \ . \label{uz2}
\end{equation}

From Eq.\ (\ref{entroq}) we finally obtain
\begin{equation*}
    \tilde S_q = \frac{h^{\nu(q-1)}}{[D(E_0)]^q}
    \int_I dE\,\omega(E)|E-E_0|^{-pq}
\end{equation*}
whence, using Eqs.\ (\ref{pq}), (\ref{di}) and (\ref{uz}),
\begin{eqnarray}\label{tildes2}
\tilde S_q &=& h^{-\nu/ p} [D(E_0)]^{1/p -1}\zeta(E_0) \nonumber \\
&=& \left|\frac{\tilde\beta}q\right| \frac{\zeta(E_0)}{D(E_0)} =
\frac{\tilde\beta}{1-q}\Theta\ .
\end{eqnarray}

The expressions obtained in this subsection will be useful in the
following of the paper, when we will have at our disposal an
explicit expression for the function $\zeta$.

\section{Power-law density-of-states functions} \label{pdof}

Let us suppose that the functions $\Omega$ of the two weakly
interacting systems ${\cal S}_1$ and ${\cal S}_2$, already
considered in Sec.\ \ref{prelim}, are given respectively by $
\Omega_1(E_1)=B_1E_1^{s_1}$ and $\Omega_2(E_2)=B_2E_2^{s_2}$, with
$s_1>0$, $s_2>0$. According to Eq.\ (\ref{omega}) the function
$\Omega$ for the system $\cal S$, composed of ${\cal S}_1$ and
${\cal S}_2$, is then given by
\begin{eqnarray*}
\Omega(E) &=&
B_1B_2s_1\int_0^E dE_1\, E_1^{s_1-1}(E-E_1)^{s_2} \nonumber \\
&=& B_1B_2s_1 E^{s_1+s_2} \int_0^1 dx\,x^{s_1-1}(1-x)^{s_2} \ .
\end{eqnarray*}
Since the integral in the last expression has the value
$\Gamma(s_1)\Gamma(s_2+1)/\Gamma(s_1+s_2+1)$ \cite{abram,almpot},
we obtain
\begin{equation}
\Omega(E)=BE^{s} \label{omegap}
\end{equation}
with
\begin{equation}
s = s_1+s_2 \label{sequil}
\end{equation}
and
\[
B=B_1B_2\frac{\Gamma(s_1+1)\Gamma(s_2+1)}{\Gamma(s_1+s_2+1)}\ .
\]
Note that the last equation can also be put in the expressive form
\begin{equation}
\log[B\Gamma(s+1)]=\log[B_1\Gamma(s_1+1)]+\log[B_2\Gamma(s_2+1)]\
. \label{bequil2}
\end{equation}
From Eq.\ (\ref{omegap}) it also follows that $\omega(E) =
BsE^{s-1}$.

Let us further assume that the system $\cal S$ is described by a
power-law probability distribution of the form (\ref{phi}). The
function $\zeta$ defined by Eq.\ (\ref{zeta}) can be explicitly
calculated as
\begin{equation}
\zeta(E_0) = Bs\int_I dE\,E^{s-1}|E-E_0|^{1-p} = A|E_0|^{s-p+1}\ ,
\label{pzeta}
\end{equation}
where \cite{abram} for $p<1$
\begin{subequations} \label{aa}
\begin{equation}
A=Bs\int_0^1 dx\,x^{s-1}(1-x)^{1-p} =
B\frac{\Gamma(s+1)\Gamma(2-p)}{\Gamma(s-p+2)}\ , \label{a1}
\end{equation}
whereas for $p>s+1$ (recall that values of $p$ between 1 and $1+s$
are forbidden)
\begin{equation}
A = Bs\int_0^{+\infty}dx\,\frac{x^{s-1}}{(x+1)^{p-1}} =
B\frac{\Gamma(s+1) \Gamma(p-s-1)}{\Gamma(p-1)}\ . \label{a2}
\end{equation}
\end{subequations}
By using Eq.\ (\ref{prob2}) we then obtain
\begin{equation}
D(E_0) = A\frac{p-s-1}{p-1}|E_0|^{s-p}\ . \label{dmeno1}
\end{equation}

We can also calculate the marginal distribution for the subsystem
${\cal S}_1$ using Eq.\ (\ref{margin1}). This gives
\[
\Phi_1(E_1)=\frac{B_2s_2}D \int_{I_1}dE_2\,E_2^{s_2-1}
|E_1+E_2-E_0|^{-p} \,
\]
where $I_1=[0,+\infty)$ for $p>1$ and $I_1=[0,E_0-E_1)$ for $p<1$.
It is then immediate to see that
\begin{subequations} \label{pmargin}
\begin{equation}
\Phi_1(E_1) = D_1^{-1}|E_1-E_0|^{-p_1}\ ,
\end{equation}
where $p_1=p-s_2$, while the coefficient $D_1$ can be expressed as
a function of $E_0$, $B_1$, $s_1$, and $p_1$ according to the same
formulae (\ref{dmeno1}) and (\ref{aa}) which give the dependence
of $D$ on $E_0$, $B$, $s$, and $p$. In the same way, from Eq.\
(\ref{margin2}) one finds
\begin{equation}
\Phi_2(E_2) = D_2^{-1}|E_2-E_0|^{-p_2} 
\end{equation} \end{subequations}
with $p_2=p-s_1$. When $p<1$ it is also obvious that
$\Phi_1(E)=\Phi_2(E)=0$ for $E>E_0$.

Equations (\ref{pmargin}) show the remarkable fact that, when the
densities of states of the systems under consideration exhibit a
power-law dependence on the energy, power-law probability
distribution functions are compatible with the condition of
thermal equilibrium which was described in Sec.\ \ref{prelim}. We
have in fact that the two weakly interacting systems ${\cal S}_1$
and ${\cal S}_2$, and the composed system $\cal S$, are all
simultaneously described by stationary power-law distributions of
the form (\ref{phi}). We note also that all these distributions
have in common the same value of $E_0$, whilst the relation among
$p_1$, $p_2$, and $p$ can be written as
\begin{equation}
r=r_1=r_2 \ , \label{erreq}
\end{equation}
where we have introduced the new parameter
\begin{equation}
r= p-s \ . \label{erre}
\end{equation}
This means that $E_0$ and $r$ are intensive quantities which, like
$\Theta$, assume the same value for two systems at mutual
equilibrium, independently of the respective size and physical
nature. From Eqs.\ (\ref{beta0z}) and (\ref{pzeta}) we
obtain
\begin{equation}
\Theta=\frac{E_0}{1-r} \ , \label{thetae0}
\end{equation}
which implies that only two of these three quantities are actually
independent. It follows that we can identify $r$ with the quantity
that was introduced as a parameter of the function $f_r$ in Eq.\
(\ref{beta}). On the other hand, Eqs.\ (\ref{sequil}) and
(\ref{bequil2}) show that the quantities $s$ and
$\log[B\Gamma(s+1)]$ are extensive, since their values for the
complete system $\cal S$ are the sum of the corresponding values
for the two subsystems ${\cal S}_1$ and ${\cal S}_2$. The same
conclusions can obviously be extended to an arbitrary number of
subsystems.

For the reasons that we have illustrated, systems possessing both
a power-law density-of-states and a power-law probability
distribution function (at least within an acceptable
approximation) are likely to be the most significant ones with
respect to their thermodynamic properties. In the following
sections we shall refer to them shortly as ``power-law systems''.

\section{A classification of power-law systems} \label{class}

Let us consider a power-law system characterized by Eqs.\
(\ref{omegap}) and (\ref{phi}). Then from Eqs.\ (\ref{uz2}) and
(\ref{thetae0}) we obtain for the mean energy
\begin{equation}
U= \frac{s}{s-p+1}E_0=s\Theta \ . \label{ue0}
\end{equation}
In order to make a comparison with the Boltzmann--Gibbs
distribution, let us rewrite the function $\Phi$ in the form
\begin{equation}
\Phi(E)= D^{-1}|E_0|^{-p}\exp\left[-p\log(1 - E/E_0) \right]\ .
\label{phiexp}
\end{equation}
For $E\ll |E_0|$ we have $\log(1-E/E_0) = -E/E_0 + O(E^2/E_0^2)$,
and the term quadratic in the energy gives a negligible
contribution to the exponential function in Eq.\ (\ref{phiexp})
whenever
\begin{equation}\label{cond1}
    |p|\frac{E^2}{E_0^2}\ll 1 \ .
\end{equation}
It is now clear that, in order to evaluate the overall behavior of
the statistical ensemble under consideration, only energies of the
same order of magnitude as $U$ need be considered, since the
probability that the system may take energy values much higher
than $U$ is in general vanishingly small. If we put $E=xU =E_0
xs/(s-p+1)$, where $x$ is a dimensionless variable, we see that
the inequality $E\ll |E_0|$ will be satisfied for all $x$ up to
order unity provided that $|p-s-1|\gg s$, whereas the condition
(\ref{cond1}) becomes equivalent to $(p-s-1)^2/|p|\gg s^2$. It is
then easy to see that these two conditions will both hold only
when $|p|\gg s^2$. Recalling Eq.\ (\ref{pq}), this is equivalent
to
\begin{equation}
|q-1|=\frac 1{|p|} \ll \frac 1 {s^2} \ .
\end{equation}
In this limit we can rewrite Eq.\ (\ref{phiexp}) as
\begin{equation}
\Phi(E)\simeq D^{-1}|E_0|^{-p}\exp(pE/E_0)\ , \label{phiexpbg}
\end{equation}
which corresponds to a canonical distribution with parameter
$\beta=-p/E_0$. We thus find again the well-known result that a
$q$-distribution tends to the Boltzmann--Gibbs distribution in the
limit $q\rightarrow 1$. In addition, the above analysis gives the
size of the neighborhood of 1 in which the parameter $q$ has to
fall in order for the two distributions to be practically
equivalent over the relevant range of energies. This size is of
order $1/s^2$, and thus decreases as the inverse square of the
dimension of the system (recall for instance that, for an ideal
monoatomic gas consisting of $N$ particles, one has $s=3N/2$).

We already know from Sec.\ \ref{defin} that for $0<p<1$ one has
$\Phi(E)=0$ for $E >E_0$ and
\[
\lim_{E\rightarrow E_0^-}\Phi(E) = +\infty \ .
\]
Recalling Eqs.\ (\ref{a1}) and (\ref{dmeno1}) we can write for
$E<E_0$
\[
\Phi(E)= \frac {\Gamma(s-p+1)} {B\Gamma(s+1)\Gamma(1-p)}
E_0^{p-s}(E_0-E)^{-p}
\]
whence
\begin{equation}
\Phi(E)\sim \frac{1-p} {BsE_0^{s-1}(E_0-E)} \rightarrow 0 \quad
\text {for} \quad p\rightarrow 1^- \ .
\end{equation}
Therefore $\Phi(E)$ vanishes in the limit $p\rightarrow 1^-$ for
any $E\neq E_0$, and the probability density on the phase space
becomes fully concentrated on the surface of equation $H(z)=E_0$.
This means that the power-law distribution tends to the \emph
{microcanonical ensemble} in the limit $p\rightarrow 1^-$, which
in turn corresponds to $q\rightarrow 0^-$. If we consider again a
system ${\cal S}$ with a subsystem ${\cal S}_1$, and we suppose
that $\cal S$ has a microcanonical distribution, so that $p\simeq
1$, then using Eq.\ (\ref{erreq}) we have $p_1=p-s+s_1<1$, since
$s>s_1$. This is in agreement with the observation, already made
some years ago \cite{plastino}, that a power-law distribution with
finite energy cutoff may arise as the marginal of a microcanonical
distribution for a finite system.

The region $s+1<p<s^2$ is the one in which power-law distributions
display features which are most remarkably different from other
known ensembles. Here the tail of the distribution at high
energies becomes relevant, and represents a significant departure
from the exponential decrease of the canonical ensemble. The
consequences of this phenomenon become extreme in the limit
$p\rightarrow (s+1)^+$, which corresponds to $q\rightarrow
[s/(s+1)]^+$. Note that for macroscopic systems, having very large
$s$, this limit value for $q$ still looks extremely close to 1,
i.e., to the value which characterizes the Boltzmann--Gibbs
distribution. Therefore $q$ does not appear to be the right
parameter to distinguish among the different possible behaviors of
a power-law system. To this purpose it is instead useful to
introduce the parameter
\begin{equation}
\eta=\frac{p-1}{p-s-1}=1+\frac s{r-1}\ .
\end{equation}
Recalling that $p$ can vary in the set $(-\infty,1)\cup
(s+1,+\infty)$, we see that $\eta$ takes values in the full
interval $(0,+\infty)$. The canonical ensemble, for which $q=1$,
$p=\pm\infty$, corresponds to $\eta=1$; the microcanonical limit
$q\rightarrow 0^-$, $p\rightarrow 1^-$ corresponds to
$\eta\rightarrow 0^+$, and the ``long power-tail'' limit
$p\rightarrow (s+1)^+$, $q\rightarrow [s/(s+1)]^+$ corresponds to
$\eta\rightarrow +\infty$. For $0<\eta<1$ the power-law
distribution is in some sense intermediate between the
microcanonical and the canonical ensembles. According to the
previous discussion, the behavior of the system becomes
essentially canonical for $|\eta-1|\ll 1/s$. A scheme of the
behavior of the parameters $p$, $q$ and $\eta$ is reported in
Table \ref{tav1}.

\begin{table*}
\caption{\label{tav1}Values of the parameters $p$, $q$ and $\eta$
for the different possible types of power-law systems. Symbols
$\nearrow$ and $\searrow$ indicate increase or decrease
respectively.}
\begin{ruledtabular}
\begin{tabular}{cccccccc}
& microcanonical & & & & canonical & & long power-tail\\
\hline $p$ & $1$ & $\searrow$ & $0$ & $\searrow$ &
$-\infty;+\infty$ &
$\searrow$ & $s+1$ \\
$q$ & $0$ & $\searrow$ & $-\infty;+\infty$ & $\searrow$ & $1$ &
$\searrow$ & $s/(s+1)$ \\
$\eta$ & $0$ & $\nearrow$ & $1/(s+1)$ & $\nearrow$ & $1$ &
$\nearrow$ & $+\infty$
\end{tabular}
\end{ruledtabular}
\end{table*}

Let us write down the relation between the parameters $\eta$ and
$\eta_1$ of two systems $\cal S$ and ${\cal S}_1$ at equilibrium
with each other. Since $r-1=s/(\eta-1)$, from Eq.\ (\ref{erreq})
we obtain
\begin{equation}
\eta_1 =1+\frac{s_1}{r_1-1}=1+s_1\frac{\eta-1}{s}\ . \label{eta}
\end{equation}
If we now consider a large power-law system $\cal S$ composed of
many weakly interacting microscopical particles, it is interesting
to take as ${\cal S}_1$ the subsystem constituted by a single
particle, for which the parameter $s_1$ is typically of order
unity (for an ideal monoatomic gas one has for instance $s=3N/2$,
$s_1=3/2$). According to the previous analysis, the one-particle
distribution will be appreciably different from Boltzmann--Gibbs
when $|\eta_1 -1|\geq 1/s_1 = O(1)$, which on account of Eq.\
(\ref{eta}) is equivalent to $|\eta-1|s_1/s\geq O(1)$. For $s\gg
s_1$ this can clearly occur only when $\eta \geq s$, which
corresponds to a system $\cal S$ far in the long power-tail
regime.

\section{Thermodynamic functions of power-law systems} \label{thermo}

\subsection{Temperature and entropy} \label{temp}

Let us now consider the crucial problem of determining the
thermodynamic temperature and entropy of a power-law system.
According to Eqs.\ (\ref{beta}) and (\ref{betaii}) we must have
\begin{equation}
\beta=f_r(\Theta)=\tilde\beta g'_q (\tilde S_q)\ . \label{beta1}
\end{equation}
Taking into account Eq.\ (\ref{tildes2}), the above equation can
be rewritten in the form
\begin{equation}
\Theta f_r(\Theta)= (1-q)\tilde S_q g'_q (\tilde S_q)\ .
\label{beta2}
\end{equation}
When we introduced the function $g_q (\tilde S_q)$ in Eq.\
(\ref{esseq}), we made the assumption that it may contain the only
parameter $q$ (see the last part of Appendix \ref{appa} for some
related considerations). We then observe that the two variables
$r$ and $\Theta$, that appear on the left hand side of Eq.\
(\ref{beta2}), are independent of the two variables $q$ and
$\tilde S_q$ that appear on the right hand side. It can in fact be
easily checked (see Appendix \ref{appa}) that all these four
quantities can be expressed as independent functions of the four
free parameters $p$, $E_0$, $s$, and $B$ that characterize the
system according to Eqs.\ (\ref{phi}) and (\ref{omegap}). It is
then obvious that the two members of Eq.\ (\ref{beta2}) can be
identically equal to each other only if they are both equal to a
common constant value $c$. If we first consider the second member,
we can thus write
\begin{equation}
g'_q (\tilde S_q)= \frac c {(1-q)\tilde S_q}\ . \label{gprime}
\end{equation}
Integrating this equation and recalling Eq.\ (\ref{esseq}), we
obtain that the entropy $S_q$ of the power-law system is given by
\begin{equation} \label{integ}
S_q=kg_q(\tilde S_q)=ck\frac{\log \tilde S_q}{1-q}+I(q)\ ,
\end{equation}
where $I(q)$ is an \emph {a priori} arbitrary function of $q$. The
requirement that the Boltzmann--Gibbs expression (\ref{sbg}) has
to be recovered in the limit $q \rightarrow 1$ imposes the
conditions $c=1$ and $I(1)=0$. If we further require that, in
accordance with the third law of thermodynamics, the entropy
vanishes at zero temperature for all systems with a nondegenerate
ground state, we must set $I(q)=0$ for any $q$. In this way we
finally obtain
\begin{equation}
S_q=k\frac{\log \tilde S_q}{1-q}\ , 
\end{equation}
which coincides with the expression of R\'enyi entropy $S_q^R$
given by Eq.\ (\ref{renyi}).

Setting also the left-hand side of Eq.\ (\ref{beta2}) equal to 1
provides the expression for the inverse absolute temperature:
\begin{equation}
\frac 1{kT} \equiv\beta=f_r(\Theta)=\frac{1}{\Theta}\ .
\label{bbeta0}
\end{equation}
We therefore find that the relation $kT=\Theta$, which had already
been proved for the Boltzmann--Gibbs statistics, is valid for all
power-law systems. From Eqs.\ (\ref{thetae0}) and (\ref{ue0}) we
also find for the mean energy $U$ of the system the value
\begin{equation}
U=\frac s \beta =skT\ . \label{ut}
\end{equation}
As in standard thermodynamics, the energy is thus directly
proportional to the absolute temperature $T$, and the specific
heat at constant volume is given by
\begin{equation}
C_V \equiv\frac{\partial U}{\partial T}=sk\ .
\end{equation}

As a verification of the results obtained in this section, we
examine in Appendix \ref{appb} the consequences of adopting,
instead of Eq.\ (\ref{renyi}), the three different entropies
mentioned at the end of Sec.\ \ref{qtherm}.

\subsection{Properties of R\'enyi entropy}

On account of the results of the previous subsection, from now on
we shall directly write $S_q^R$ in place of $S_q$. From Eqs.\
(\ref{tildes2}), (\ref{di}), (\ref{dmeno1}) and (\ref{thetae0}) we
obtain
\begin{eqnarray}
k^{-1}S_q^R &=&-\nu\log h+\log A +(p-1)\log \frac{p-1}{p-s-1} \nonumber \\
&& +\ s\log(|p-s-1|kT)\ . \label{pentropy}
\end{eqnarray}
Having in mind that the number of degrees of freedom $\nu$ and the
value of the external quantities such as the volume are directly
related to the parameters $s$ and $B$ of the density of states
(\ref{omegap}), and recalling Eq.\ (\ref{pq}), we want here to
analyze the behavior of the entropy considered as a function of
the independent variables $q$, $s$, $B$, and $T$.

Let us first of all compare Eq.\ (\ref{pentropy}) with the
corresponding expression for the Boltzmann--Gibbs entropy. By
substituting for $\rho= \Phi(E)$ the expression (\ref{phibg}) into
Eq.\ (\ref{essebg}), and observing that, on account of Eq.\
(\ref{omegap}),
\begin{eqnarray*}
Z&=&\int_\Gamma dz\,e^{-\beta H}=\int_0^{+\infty} dE\,sB E^{s-1}
e^{-\beta E} \\
&=& \beta^{-s}B\Gamma(s+1) \ ,
\end{eqnarray*}
one obtains
\begin{equation}
k^{-1}S_{BG}(s,B,T) =-\nu\log h +\log[B\Gamma(s+1)]+s\log kT + s \
. \label{esse1}
\end{equation}
We can write in general
\begin{equation}
S^R_q(s,B,T)= S_{BG}(s,B,T)+\Delta S_q(s) \ , \label{srq}
\end{equation}
where, using for $A$ the expressions (\ref{aa}), we have
\begin{subequations} \label{delta}
\begin{eqnarray}
k^{-1}\Delta S_q(s)&=& \log\frac{\Gamma(1-p)}{\Gamma(s+1-p)}
+p\log(1-p) \nonumber \\
&& +\ (s-p)\log(s+1-p) -s \label{deltas}
\end{eqnarray}
for $p<1$, and
\begin{eqnarray}
k^{-1}\Delta S_q(s)&=& \log\frac{\Gamma(p-s)}{\Gamma(p)}+p
\log(p-1) \nonumber \\
&& -\ (p-s)\log(p-s-1)-s \label{deltas1}
\end{eqnarray}
\end{subequations}
for $p>s+1$. By calculating the limit of the expression
(\ref{deltas}) for $p\rightarrow -\infty$ and the limit of
(\ref{deltas1}) for $p\rightarrow +\infty$ one can easily check
that $\lim_{q\rightarrow 1}\Delta S_q(s)=0$. We can therefore
conclude that, as expected,
\[
\lim_{q\rightarrow 1}S_q^R(s,B,T)=S_{BG}(s,B,T) \ .
\]

Another interesting case to consider is the limit $q\rightarrow
0^-$, or $p\rightarrow 1^-$, which according to our discussion of
Sec.\ \ref{class} corresponds to a microcanonical distribution.
Taking into account Eq.\ (\ref{ut}) we obtain from Eqs.\
(\ref{esse1}) and (\ref{deltas})
\begin{eqnarray}
\lim_{q\rightarrow 0}S_q^R(s,B,T) &=&k(-\nu\log h +\log
B+s\log U) \nonumber \\
&=& k\log \frac {\Omega(U)} {h^\nu} \ , \label{micro}
\end{eqnarray}
which is indeed a well-known expression for the entropy in the
microcanonical ensemble \cite{gibbs,munster}.

Since the expression (\ref{esse1}) is manifestly extensive, in
order to investigate the thermodynamic limit it is only necessary
to study the behavior of $\Delta S_q$ for $s\rightarrow +\infty$.
If we treat $q$ (and therefore also $p$) as a constant, recalling
that $p$ can never lie in the interval $[1,s+1]$ we see that such
a limit can only be taken when $p<1$. Therefore, applying
Stirling's formula to the expression (\ref{deltas}) we obtain
\begin{equation}
\Delta S_q(s)\sim -\frac 3 2 k\log s \quad \text {for} \quad
s\rightarrow +\infty \ . \label{limd}
\end{equation}

From a physical point of view it might seem more appropriate to
consider a thermodynamic limit in which the intensive quantity $r$
is kept constant instead of $q$. We have in this case $p=s+r
\rightarrow +\infty$, which implies $p>s+1$. Then from Eq.\
(\ref{deltas1}) we obtain
\begin{equation}
\Delta S_q(s)\sim -\frac 12 k\log s \quad \text {for} \quad
s\rightarrow +\infty \label{limd2}
\end{equation}
with $r=$ constant, $q=1-1/(s+r)\rightarrow 1^-$.

Let us finally consider the thermodynamic limit in which we keep
constant the parameter $\eta$ introduced in Sec.\ \ref{class}. We
have in this case to substitute $p=\eta s/(\eta -1)+1$ into Eq.\
(\ref{deltas}) when $0<\eta<1$, or into Eq.\ (\ref{deltas1}) when
$\eta>1$, and then to take the limit for $s\rightarrow +\infty$.
We obtain in this way
\begin{equation} \label{limd3}
\lim_{s\rightarrow +\infty}\Delta S_q(s)=\begin{cases} \frac 3 2
k\log\eta \quad &\text{for} \quad
0<\eta<1 \\
-\frac 1 2 k\log\eta\quad &\text{for}\quad \eta\geq 1 \end{cases}
\end{equation}
with $\eta=$ constant, $q=s/(s+1- \eta^{-1}) \rightarrow 1$.

We see that in the limits considered $\Delta S_q$ either increases
logarithmically with the size of the system, as in Eqs.\
(\ref{limd})--(\ref{limd2}), or tends to a constant value as in
Eq.\ (\ref{limd3}). It follows that in all cases the extensive
part $S_{BG}$ prevails in Eq.\ (\ref{srq}), and the entropy
$S_q^R(s,B,T)$ essentially reduces to $S_{BG}(s,B,T)$ as
$s\rightarrow +\infty$. We note also that $q$ does not appear in
the expression (\ref{ut}) of the mean energy. These results can be
seen as a generalization of the well-known theorem about the
equivalence of the different ensembles in the thermodynamic limit.
We have shown in fact that this equivalence does not only hold
between the canonical and the microcanonical ensembles, but also
among the power-law distributions corresponding to all possible
values of $\eta$. This fact is all the more remarkable if one
recalls that, as we have illustrated in detail in Sec.\
\ref{class}, the microscopical distribution function can instead
be significantly different from Boltzmann--Gibbs also for very
large systems.

We would like finally to mention that the additivity of R\'enyi
entropy for mutually independent systems was already well known.
Our analysis has shown on the other hand that it is also additive
for large power-law systems which, instead of being independent,
are at mutual thermal equilibrium in the sense that we have
specified in Sec.\ \ref{prelim}. One can observe nevertheless that
the R\'enyi entropy (\ref{srq}) includes, at variance with the
Boltzmann--Gibbs one, a nonextensive contribution $\Delta S_q$
which may be nonnegligible for systems of small size.

\subsection{The ideal gas} \label{ideal}

Let us consider as a significant example the case of an ideal
monoatomic gas composed of $N$ particles of mass $m$ in a volume
$V$. We have
\begin{eqnarray*}
\nu &=& 3N \\
s &=& \frac {3N}2 \\
B &=& \frac {(2\pi m)^{3N/2}V^N} {\Gamma(3N/2+1)N!}\ ,
\end{eqnarray*}
where the factor $N!$ in the denominator of the last expression
accounts for the indistinguishability of the atoms. From Eqs.\
(\ref{ut}) and (\ref{esse1}) we then obtain for large $N$
\begin{eqnarray}
U &=& \frac 3 2 NkT  \label{ukt} \\
S^R_q &=& Nk\left[\frac 3 2\log\frac{2\pi mkT} {h^2} +\log \frac V
N +\frac 5 2\right] \nonumber \\
&&+\ O(\log N) \ . \label{sideal}
\end{eqnarray}
We find in this way two formulae which are already familiar from
standard statistical thermodynamics. It is then easy to see that
ideal gases governed by power-law distributions obey the same
state equation of Boltzmann--Gibbs ideal gases. Consider in fact
an infinitesimal isothermal transformation, for which $dT=0$. Then
from Eq.\ (\ref{ukt}) we get $dU=0$, and the infinitesimal work
$\delta W =PdV$ performed by the system, where $P$ is the pressure
of the gas, can be obtained according to the first and second laws
of thermodynamics as
\begin{equation}
PdV=\delta Q=TdS_q^R=NkT\frac{dV}{V} \ ,
\end{equation}
where the last equality follows from Eq.\ (\ref{sideal}).
Comparing the first and last members of the above equation we
finally get
\begin{equation}
PV=NkT \ . \label{idgas}
\end{equation}
In particular, one can conclude that also for power-law systems
the ideal gas temperature defined according to Eq.\ (\ref{idgas})
is the same as the thermodynamic temperature based on the second
law.

\section{Conclusion}

We have developed a new generalized thermodynamics for systems
governed by Tsallis distributions, in a way which also allows for
a rigorous statistical interpretation of the zeroth law. Our
approach thus provides a new point of view on an issue which has
been the object of sharp debate in recent literature \cite{nauen}.
The main outcome of our analysis is that the thermodynamic entropy
associated with these distributions is expressible as a function
of the probabilities according to the formula first introduced by
R\'enyi.

Our results have been obtained under a minimum set of simple and
apparently plausible hypotheses. Probably the main new ingredient,
with respect to former investigations, is the observation that
Tsallis statistics is capable of describing stationary states, in
which power-law systems combine together in such a way to give
rise to a larger composed power-law system. The consideration of
such a circumstance allows one to extend to a deeper level the
analogy with the standard thermodynamics of weakly interacting
systems \cite{gibbs,khin}, which has already proved to be an
illuminating guideline for the analysis of Tsallis
thermostatistics \cite{plastino,almeida1,adib,
almpot,potiguar,almeida}.

Of course power-law distributions can in principle apply also to
systems for which the hypotheses of weak coupling or of power-law
density of states are in general not satisfied. An important case
of this type is likely to be represented by systems of particles
with long-range interactions, which are indeed considered among
the most interesting candidates for the applicability of
nonextensive thermodynamics. In such cases the conclusions of our
analysis may not necessarily hold, and it is possible that other
theoretical investigations are to be carried out on the basis of a
completely different approach. In particular, other descriptions
might become appropriate, in which no reference at all is made to
the usual formalism of equilibrium thermodynamics. For instance, a
justification for Tsallis entropy has recently been proposed on
the basis of purely dynamical assumptions about the behavior of a
system far from equilibrium \cite{carati1}. It has also been
pointed out that deviations from the exact Tsallis distribution
law can be present in real situations \cite{beck}. It is
reasonable to expect that such deviations may be related to the
specific form of the density-of-states function, so that a
generalization of Tsallis statistics might be required in order to
treat the cases in which the density of states significantly
departs from a power law.

\begin{acknowledgments}
The author wishes to thank A.\ Carati, L.\ Galgani and A.\ Ponno
for interesting discussions.
\end{acknowledgments}

\appendix

\section{On the independent parameters of a power-law system}
\label{appa}

The deduction of Eq.\ (\ref{gprime}) from Eq.\ (\ref{beta2}) is
based on the mutual independence of the four variables $r$,
$\Theta$, $q$ and $\tilde S_q$. In order to elucidate this point,
let us start from the four main free parameters which can be put
at the basis of the description of a power-law system. These are
$s$, $B$, $p$ and $E_0$. The first two of them appear in the
expression (\ref{omegap}) which gives the primitive function of
the density of states, and are therefore directly related to the
form of the hamiltonian function. Their physical meaning is
illustrated by the example of the ideal gas which is treated in
Sec.\ \ref{ideal}. In particular, $2s$ expresses the number of
degrees of freedom contributing to the hamiltonian, and is
therefore related to the size (number of particles) of the system,
while $B$ contains the dependence on the external parameters
which, like the volume $V$, are varied when the system exchanges
mechanical work with the surrounding environment. On the other
hand, $p$ and $E_0$ are the two free parameters in the power-law
probability distribution (\ref{phi}), $D$ being a normalization
constant. According to Eqs.\ (\ref{pq}) and (\ref{ue0}), $p$ is
simply a function of $q$, while $E_0$ fixes the mean energy, and
thus the temperature of the distribution. For the case of the
ideal gas, for instance, the mutual independence of $s$, $B$,
$E_0$ and $p$ amounts to the mutual independence of $N$, $V$, $T$
and $q$.

Assuming for simplicity the relation $\nu =2s$ as for the ideal
gas (the particular relation between these two parameters is
inessential with respect to the present considerations), we can
express $q$, $r$, $\Theta$ and $\tilde S_q$ as functions of the
independent quantities $s$, $B$, $E_0$ and $p$, according to the
relations
\begin{eqnarray}
q &=& 1-\frac 1p \label{uno}\\
r &=& p-s \label{due}\\
\Theta &=&\frac{E_0}{1+s-p} \label{tre}\\
\tilde S_q &=&\frac{1-p}{s+1-p}\left(BG(s,p)\Gamma(s+1)
\frac{|E_0|^s}{h^{2s}}\right )^{1/p} \label{quattro}
\end{eqnarray}
Equations (\ref{uno})--(\ref{tre}) are identical respectively to
Eqs.\ (\ref{pq}), (\ref{erre}) and (\ref{thetae0}). Equation
(\ref{quattro}), in which we have introduced the function
\begin{equation*}
G(s,p)\equiv \begin{cases}\Gamma(1-p)/\Gamma(s-p+1)\quad
& \text{for} \quad p < 1 \\
\Gamma(p-s)/\Gamma(p)\quad & \text{for} \quad p > s+1 \ ,
\end{cases}\end{equation*}
is derived from Eqs.\ (\ref{tildes2}) and
(\ref{pzeta})--(\ref{dmeno1}).

It easy to see that the four functions defined by Eqs.\
(\ref{uno})--(\ref{quattro}) are mutually independent. They can in
fact be explicitly inverted, so as to express $s$, $p$, $E_0$ and
$B$ as functions of $q$, $r$, $\Theta$ and $\tilde S_q$. We obtain
\begin{eqnarray}
s &=& \frac{1}{1-q} -r \label{cinque}\\
p &=& \frac{1}{1-q} \label{sei}\\
E_0 &=& (1-r)\Theta \\
B &=& \frac{K(q,r)} {\Gamma (1/(1-q)-r+1)} \nonumber \\
&&\times\left( \frac{(q-1)(1-r)}{q}\tilde S_q \right)^{1/(1-q)}
\nonumber \\
&&\times\left( \frac{h^2}{\Theta|1-r|}\right)^{1/(1-q)-r} \ ,
\label{otto}
\end{eqnarray}
where
\begin{equation*}
K(q,r)\equiv \begin{cases}
\displaystyle{\frac{\Gamma(1-r)}{\Gamma(q/(q-1))}} \quad &
\text{for} \quad q<0,\ q>1\\[8pt]
\displaystyle{\frac{\Gamma(1/(1-q))}{\Gamma(r)}} \quad &
\text{for} \quad \displaystyle{\frac s{s+1}}<q<1 \ .
\end{cases}\end{equation*}

The argument we have used after Eq.\ (\ref{beta2}), in order to
justify the introduction of the constant $c$, is also based on the
assumption that the function $g_q(\tilde S_q)$ does not contain
any parameter other than $q$. Note that all the functions
appearing on the right-hand sides of Eqs.\
(\ref{tsallis})--(\ref{escort}) do agree with this general
assumption, which is dictated by obvious reasons of simplicity and
economy. We expect in fact that such a fundamental physical
quantity as the entropy should have the simplest analytical
expression which is compatible with the laws of thermodynamics,
without the inclusion of unnecessary parameters. The argument
about the constancy of $c$ is also at the basis of the remarkable
equality, which we have found in Sec. \ref{temp}, between the
``empirical temperature'' $\Theta$ and the thermodynamic
temperature $kT$. It is however interesting to examine the
consequences of restraining from the above assumption, and
considering in Eq.\ (\ref{esseq}) a function $g_{q,r}(\tilde S_q)$
which may also depends on the parameter $r$. This would imply the
replacement of the constant $c$ on the right-hand side of Eq.\
(\ref{integ}) with a function $c(r)$, which would then remain as a
multiplicative coefficient also in the final expressions of the
entropy and of the inverse temperature:
\[ 
S_q =kc(r)\frac{\log \tilde S_q}{1-q}\ , \qquad\qquad T =\frac
\Theta{kc(r)} \ .
\] 
Taking into account Eqs.\ (\ref{due}) and (\ref{sei}), one should
then be forced to require that $\lim_{r\to \pm\infty}c(r)=1$, in
order for the Boltzmann--Gibbs entropy and temperature to be
recovered in the limit $q\to 1$. Furthermore, we have shown in
Sec.\ \ref{class} that a power-law distribution tends to the
microcanonical ensemble for $q\to 0$. If one then requires that
the correct microcanonical entropy (\ref{micro}) and the empirical
temperature $kT=\Theta= U/s$ \cite{munster} be obtained for any
$s>0$ in the limit $q\to 0$, it would also be necessary to impose
$c(r)= 1$ for all $r$ such that $-\infty <r< 1$. These can be
considered as additional arguments in favor of the natural choice
of taking $c=1$ for all $r$.

\section{Comparison among alternative entropies} \label{appb}

In Sec.\ \ref{temp} we have proved that the condition of
compatibility with the zeroth law of thermodynamics univocally
determines the expression of the entropy for a power-law system.
As a verification of this result, it is interesting to check
directly how this condition of compatibility is violated by the
three particular functions which have been mentioned at the end of
Sec.\ \ref{qtherm} as well-known possible alternatives to R\'enyi
entropy.

It follows from the first and second laws of thermodynamics that
the inverse of the thermodynamic temperature $\beta$ must be
related to the entropy $S$ by the equation
\begin{equation}
k \beta =\frac{\partial S}{\partial U} \ ,
\end{equation}
where the partial derivative of the entropy is evaluated at
constant external parameters (i.e., those parameters, such as the
volume of the system, whose variation is related to the production
of mechanical work). When the entropy has the form (\ref{esseq})
we have then
\begin{equation} \label{appbeta}
\beta = g'_q(\tilde S_q)\frac{\partial \tilde S_q}{\partial U} \ .
\end{equation}
From Eqs.\ (\ref{tildes2}), (\ref{pzeta}), (\ref{dmeno1}) and
(\ref{ue0}) we obtain
\begin{equation*}
\tilde S_q=\left[h^{-\nu}A\left(\frac{p-1}{p-s-1}\right)^{p-1}
\left( \frac{|p-s-1|}s U\right)^s\right]^{1/p}\ . 
\end{equation*}
According to Eq.\ (\ref{aa}), the quantity $A$ depends on the
external parameters (through $B$) but not on the internal energy
$U$. We therefore obtain
\begin{equation*}
\frac{\partial \tilde S_q}{\partial U} = \frac{s \tilde S_q}{p U}
=\frac {\tilde S_q}{p\Theta}\ .
\end{equation*}
Substituting for $g'_q$ in Eq.\ (\ref{appbeta}) the derivatives of
the expressions which appear on the right-hand sides of Eqs.\
(\ref{tsallis})--(\ref{escort}), we then find for the function
$\beta$ in the four respective cases
\begin{eqnarray*}
\beta^T &=& \frac {\tilde S_q}\Theta \\
\beta^R &=& \frac {1}\Theta \\
\beta^N &=& \frac 1{\tilde S_q\Theta} \\
\beta^E &=& \frac 1{\tilde S_q^{1/q}\Theta}\ .
\end{eqnarray*}
We thus see that only $\beta^R$, derived from R\'enyi entropy, is
equal to the inverse of the parameter $\Theta$ which was
introduced in Sec.\ \ref{prelim} on the basis of the zeroth law of
thermodynamics. On the contrary, $\beta^T$, $\beta^N$ and
$\beta^S$, which are respectively associated with the entropies
(\ref{tsallis}), (\ref{normal}) and (\ref{escort}), depend also on
$\tilde S_q$, which in general takes different values for systems
at thermal equilibrium with each other. It is therefore confirmed,
by the analysis of these examples, that only R\'enyi entropy leads
to an expression for the thermodynamical temperature which is
compatible with the zeroth law. This result obviously does not
exclude that Tsallis or other entropies can nevertheless be
conveniently employed whenever some of the starting hypotheses of
the present work (such as weak coupling or power-law density of
states) are manifestly violated by the system under investigation.
In such circumstances a proper modification of the standard laws
of thermodynamics will also be necessary, as it has already been
recognized in the literature.

\end{document}